\documentclass[english,a4paper,12pt]{article}
\usepackage[T2A]{fontenc}
\usepackage[cp1251]{inputenc}
\usepackage{babel}
\usepackage{amsmath}
\usepackage{graphicx}
\usepackage{amssymb}
\usepackage{epsf}

\newcommand{\veq}[2]{\upsilon_{#1 #2}}
\newcommand{\vn}[1]{\upsilon_{#1}}
\newcommand{\veqt}[2]{\overline{\upsilon}_{#1 #2}}
\newcommand{\WID}[2]{#1_{\mathrm{#2}}}
\newcommand{\mut}[1]{\widetilde{\mu}_{#1}}
\newcommand{\vnt}[1]{\widehat{\upsilon}_{#1}}
\newcommand{\psn}[1]{\psi_{#1}}
\newcommand{\id}{\mathrm{id}}
\newcommand{\Dif}[3]{\!\left(\!\frac{\partial #1}{\partial #2}\!\right)_{\!\!
#3}\!}
\newcommand{\Avsg}[1]{\langle\sigma^{#1}\rangle}
\newcommand{\sgt}[2]{\widehat{\sigma}_{#1}^{#2}}
\newcommand{\gccm}{\mbox{g$\,$cm$^{-3}$}}

\newcommand{\rgccm}{\mbox{g\,cm$^{-3}$}}
\newcommand{\kbz}{k_{\mathrm{b}}}
\newcommand{\Ye}{\mbox{$Y_{\mathrm e}$}}
\newcommand{\Yi}{\mbox{$Y_{\mathrm i}$}}
\newcommand{\nucmu}{\mbox{$m_{\mathrm u}$}}

\begin{document}

\title{Excluded-Volume Approximation for Supernova Matter}
\author{A.V. Yudin\thanks{yudin@itep.ru}}
\date{}
 \maketitle
\begin{center}
 {\em Institute for Theoretical and Experimental Physics,
ul. Bol’shaya Cheremushkinskaya 25, Moscow, 117259 Russia}
\end{center}
\begin{abstract}
A general scheme of the excluded-volume approximation as applied
to multicomponent systems with an arbitrary degree of degeneracy
has been developed. This scheme also admits an allowance for
additional interactions between the components of a system. A
specific form of the excluded-volume approximation for
investigating supernova matter at subnuclear densities has been
found from comparison with the hard-sphere model. The possibility
of describing the phase transition to uniform nuclear matter in
terms of the formalism under consideration is discussed.
\end{abstract}

\vspace*{0.3cm}

\thispagestyle{empty} \clearpage

\section*{INTRODUCTION}
The evolution of massive stars ends with the formation of a
central ``iron'' core. This core loses its hydrodynamic stability
and is drawn into gravitational collapse that eventually leads to
a supernova explosion. In the course of collapse, the density in
the central regions of the star rises to values comparable to the
density of matter inside atomic nuclei, $\WID{\rho}{n}\simeq
2.6\times 10^{14}\ \rgccm$. Under these conditions, when the mean
distances between atomic nuclei are comparable to their sizes, a
strong nuclear interaction confining the nucleons inside the
nucleus begins to manifest itself. Nuclei with a mass number up to
$A\sim 100$ begin to fuse, stretch, and deform. As a result,
peculiar configurations known as ``lasagna'', ``pasta'', etc. are
formed. At a density $\rho\sim 0.5\WID{\rho}{n}$ the phase
transition to uniform nuclear matter occurs.

\begin{figure}[htb]
\begin{center}
\vspace*{-0.1cm}
\includegraphics[totalheight=8cm]{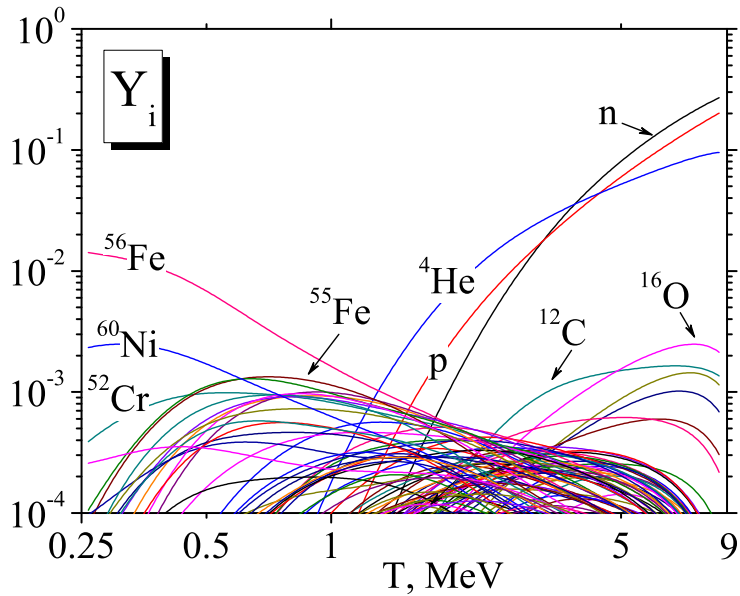}
 \vspace*{-0.2cm}
 \caption{(Color online) Chemical composition of matter at $\rho=10^{13}~\rgccm$ and $\Ye=26/56$}
\label{PicYi_Id_Gas}
 \vspace*{-0.2cm}
\end{center}
\end{figure}
Figure $\ref{PicYi_Id_Gas}$ shows an example of the calculation of
equilibrium mass fractions for matter $\Yi\equiv n_i\nucmu/\rho$,
where $n_i$ --- is the number density of component $i$, as a
function of the temperature at density $\rho=10^{13}~\rgccm$ and
leptonic number $\Ye=26/56$. The calculation was performed by
assuming the nuclear statistical equilibrium (NSE, for more
detail, see Nadyozhin and Yudin 2004) conditions to be met. As we
see, the matter consists of a mixture of free nucleons, light
(mainly helium) nuclei, and heavy iron-peak nuclei even at
subnuclear densities. The proper description of this complex
system is nontrivial and is especially important for questions
related to the chemical composition of matter: nucleosynthesis,
the problem of neutron-rich nuclei, etc.

\begin{figure}[htb]
\begin{center}
\vspace*{-0.1cm}
\includegraphics[totalheight=7cm]{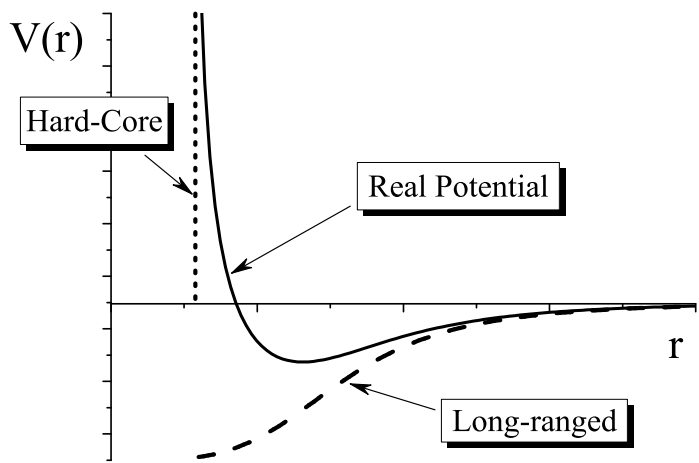}
 \vspace*{-0.2cm}
 \caption{Interaction potentials.}
\label{PicPotentials} \vspace*{-0.2cm}
\end{center}
\end{figure}

To derive a realistic equation of state capable of describing
matter in this density range and to provide a smooth transition to
uniform nuclear matter, it is necessary to introduce the
nucleon–nucleon interaction potential into the scheme for
calculating the equation of state at $\rho\leq\WID{\rho}{n}$. A
schematic view of this potential is indicated in
Fig.~\ref{PicPotentials} by the solid line.
It consists of the long--range component causing attraction and
the part leading to strong repulsion at short distances. Such a
form of the potential allows the system of nucleons to be modeled
with a good accuracy by a set of hard spheres with a certain
attractive potential between them, i.e., by a combination of the
potentials indicated in Fig.~\ref{PicPotentials} by the vertical
dotted and dashed lines. The presence of a hard--core component in
nucleons determines the dependence $R\sim A^{1/3}$ for the radius
of nuclei, where $A$ is the nuclear mass number. All these
considerations lead to the idea of describing matter in the
subnuclear range in terms of the excluded--volume approximation
(EVA) --- a well--known approach in the thermodynamics of gases
that allows for the finite size of the system’s components. In the
succeeding sections, we will attempt to construct an EVA model
that would be thermodynamically self--consistent, would naturally
describe systems with a large number of components with different
sizes, would take into account the component degeneracy effects,
and would admit the inclusion of additional (apart from the
hard--core one) interactions between particles.

\section*{BOLTZMANN GAS}
To derive the expressions for the thermodynamic quantities of a
Boltzmann gas in the EVA model, let us consider the standard
procedure (see, e.g., Landau and Lifshitz 1976) of the particle
distribution over the phase space. The system under consideration
consists of $M$ types of particles with the total number of
particles of each type $N_n,\ n=1{\div}M$. Each type of particles
is divided into $L_n$ groups, $N_n=\sum_{k=1}^{L_n}N_{n}^{k}$,
belonging to different regions of the momentum space $G_n^k=g_{n}
d^{3}p_{k}/h^{3}$, where $g_{n}=(2j_{n}{+}1)$ is the degeneracy
factor. Let us calculate the total number of states available to
such a system. For the first particle of the first type from the
first group, $G_{1}^{1}V$ states are available, where $V$ is the
system’s volume. For the second particle, the number of states is
$G_{1}^{1}(V-\upsilon_{11})$. Here,
$\upsilon_{ij}\equiv\frac{\pi}{6}(\sigma_i{+}\sigma_j)^{3}$ is the
so--called volume of the shielding sphere and $\sigma_i$ is the
diameter of the corresponding type of particles. Continuing this
procedure, we will obtain the number of states available to the
particles of the first type from the first group,
$\triangle\Gamma_{1}^{1}$:
\begin{equation}
\triangle\Gamma_{1}^{1}=\frac{G_{1}^{1}V\times
G_{1}^{1}(V{-}\upsilon_{11})\times\ldots\times
G_{1}^{1}(V{-}(N_{1}^{1}{-}1)\upsilon_{11})}{N_{1}^{1}!},
\end{equation}
where the factorial in the denominator emerges, because the
particles are identical. The number of available states is
$G_{1}^{2}(V{-}\upsilon_{11}N_{1}^{1})$ for the first particle
from the second group,
$G_{1}^{2}(V{-}\upsilon_{11}N_{1}^{1}{-}\upsilon_{11})$ for the
second particle, and so on. In general, the following formula is
valid for the number of states:
\begin{gather}
\triangle\Gamma_n^k=\frac{G_n^k(V{-}A_n^k){\times}
G_n^k(V{-}A_n^k{-}\upsilon_{nn}){\times}{\ldots} {\times}
G_n^k(V{-}A_n^k{-}\upsilon_{nn}(N_n^k{-}1))}{N_n^k!},\\
A_n^k=\sum\limits_{i=1}^{n-1}\upsilon_{ni}N_i+
\sum\limits_{j=1}^{k-1}\upsilon_{nn}N_n^j.
\end{gather}
The total number of states for the system is the product
$\triangle\Gamma=\prod_{n,k}\triangle\Gamma_{n}^{k}$ and the
entropy is defined via its logarithm:
$S=\kbz\ln\triangle\Gamma=\kbz\sum_{n,k}\ln\triangle\Gamma_{n}^{k}$,
where $\kbz$ is the Boltzmann constant. Using Stirling’s formula,
the expression for the logarithm of the number of states can be
brought to the form
\begin{equation}
\ln\triangle\Gamma_n^k=\frac{V{-}A_n^k}{\upsilon_{nn}} \ln
G_n^k(V{-}A_n^k)-\frac{V{-}A_n^{k+1}}{\upsilon_{nn}} \ln
G_n^k(V{-}A_n^{k+1})-N_n^k\ln N_n^k.
\end{equation}
Since the approximation considered here is valid only if the
excluded volume is small, the derived expression should be
expanded to give
\begin{equation}
S=\kbz\sum\limits_{n,k}N_n^k\ln\frac{G_n^k Ve}{N_n^k}-
\frac{\kbz}{2V}\sum\limits_{i,j}\upsilon_{ij}N_i N_j.
\label{S-Bolzmann}
\end{equation}
Here, the first term is the ordinary Boltzmann expression for
entropy and the second term is attributable to the
excluded--volume effect. In equilibrium, the entropy must have a
maximum at fixed values of the numbers of particles of each type
$N_n=\sum_{k}N_n^k$ and the total energy of the system
$E=\sum_{n,k}N_n^k\varepsilon_n^k$, where $\varepsilon_n^k$ is the
energy of the particles of type $n$ belonging to the $k$-th group.
The standard Lagrange multiplier method of searching for a minimum
by varying $N_n^k$ leads to the equation
\begin{equation}
\kbz\ln\frac{G_n^k V}{N_n^k}-\frac{\kbz}{V}\sum\limits_{i}
\upsilon_{ni}N_i+\alpha_n+\beta\varepsilon_n^k=0.
\end{equation}
The Lagrange multipliers are nothing but $\alpha_n=\mu_n/T$ and
$\beta=-1/T$, where $\mu_n$ is the chemical potential for the
particles of type $n$ and $T$ is the temperature. Now, we can
ultimately write
\begin{equation}
N_n^k=G_n^k\Bigl(V{-}\sum_{i} \veqt{i}{n}N_i\Bigr)\exp\biggl(
-\frac{\varepsilon_n^k}{\kbz T}+\frac{\mu_n}{\kbz
T}-\frac{1}{V}\sum_{i}\veqt{n}{i}N_i\biggr). \label{N-Bolzmann}
\end{equation}
Here, we introduced the quantities $\veqt{i}{j}$ satisfying the
condition $\veqt{i}{j}+\veqt{j}{i}=\veq{i}{j}$. In the linear (in
excluded volume) approximation, it does not matter whether we
write the correction determined by it as a factor in front of the
exponential in Eq.~$(\ref{N-Bolzmann})$ or as an addition to the
chemical potential. However, the need for expanding the domain of
applicability and comparison with other approaches force us to
write this expression precisely in this form. For identical
particles, $\veqt{i}{i}=\veq{i}{i}/2$, which corresponds to the
well-known approximation of an effective excluded volume equal to
quadruple the intrinsic particle volume. In addition, there exist
other constraints on the choice of $\veqt{i}{j}$ to be discussed
in detail below. However, certain freedom in choosing the
expression for $\veqt{i}{j}$ remains. Summing
Eq.~$(\ref{N-Bolzmann})$ over $k$, we will obtain the system of
equations relating the numbers of particles to the temperature and
chemical potentials. The sum $\sum_{k}\varepsilon_n^k N_n^k$
defines the energy of the $n$-th type of particles. Substituting
the expression for $N_n^k$ into Eq.~$(\ref{S-Bolzmann})$ give the
expression for entropy and so on.

\section*{THE GENERAL CASE} The simple scheme of allowance for the finite size
of particles described in the preceding section is unsatisfactory
for several reasons. First, it was formulated only for Boltzmann
statistics. Second, it is unclear how to take into account other
interactions that are not described by the EVA in it. And, most
importantly, it becomes inapplicable even at $V=N_n\veqt{n}{n}$
(we consider one type of particles). Let us introduce the concept
of a packing fraction $\eta=\sum_i\textbf{v}_{i}n_i$, where
$\textbf{v}_{i}$ is the intrinsic volume of the particles and
$n_{i}$ is their number density. Obviously, this quantity is the
ratio of the volume occupied by the particles to the total volume.
The equations from the preceding section are then seen to become
inapplicable at $\eta=0.25$. Meanwhile, it is well known that the
closest packing of spheres with the same size is achieved in a
face-centered cubic lattice at $\eta=\pi/3\sqrt{2}\approx 0.74$.
Moreover, numerical simulations show that the phase transition to
crystalline order in a system of hard spheres occurs at
$\eta\approx 0.6$. The cause of such an overestimation of the
excluded-volume effect in the above analysis is clear: we assumed
that each particle reduced the space available to other particles
by a value equal to the volume of the shielding sphere. For
identical particles, this is quadruple the particle volume: the
excluded volume for two particles is $\frac{\pi}{6}(2\sigma_n)^3$.
Meanwhile, this is true only in the limit of low number densities:
the effective volume per particle must decrease as they increase,
i.e., the excluded volume must be a function of the number
densities of the system’s components. Below, we will attempt to
construct a phenomenological theory that would take these
requirements into account and would be free from these
shortcomings.

We will assume that all particles are fermions. For
subnuclear-density supernova matter of interest to us, this is
always the case: the free neutrons and protons are described by
Fermi statistics, while the nuclei are far from degeneracy and are
a Boltzmann gas (the limiting case of Fermi statistics). The
expression for the logarithm of the number of states for an ideal
Fermi gas is (the notation is the same as that used above)
\begin{equation}
\ln\triangle\Gamma_n^k=G_n^k V\ln G_n^k V-
\left[G_n^kV-N_n^k\right]\ln (G_n^kV-N_n^k)- N_n^k\ln N_n^k.
\end{equation}
The substitution that we make to take into account the
excluded-volume effect is natural: $V\rightarrow V{-}\vn{n}$, with
$\vn{n}=\vn{n}(N_{1},N_{2},\ldots,N_{M})$ being a function of the
numbers of particles for all components of the system. To begin
with, let us calculate the necessary derivatives:
\begin{align}
\Dif{\ln\triangle\Gamma_n^k}{N_n^k}{\vn{n}}&=
\ln\left[G_n^k(V{-}\vn{n}){-}N_n^k\right]-\ln N_n^k,\\
\Dif{\ln\triangle\Gamma_n^k}{\vn{n}}{N_n^k}&= -G_n^k\ln
G_n^k(V{-}\vn{n})+G_n^k \ln\left[G_n^k(V{-}\vn{n}){-}N_n^k\right].
\end{align}
The derivative of the entropy with respect to $N_n^k$ is
\begin{equation}
\quad\quad\quad\Dif{S}{N_{n}^{k}}{}=\kbz\Dif{\ln\triangle
\Gamma_n^k}{N_n^k}{\vn{n}}\!+\kbz\!\sum_{p,j}
\Dif{\ln\triangle\Gamma_p^j}{\vn{p}}{N_p^j} \Dif{\vn{p}}{N_n}{}.
\end{equation}
The variational principle $\delta(S+\sum_{n}\alpha_n N_n+\beta
E)=0$ gives
\begin{multline}
\kbz\ln\left[\frac{G_n^k(V{-}\vn{n}){-}N_n^k}{N_n^k}\right]+\alpha_{n}+\beta\varepsilon_n^k+\\
+\kbz\sum_{p}\Dif{\vn{p}}{N_n}{}\sum_{i}
G_p^i\ln\left[\frac{G_p^i(V{-}\vn{p}){-}N_p^i}{G_p^i(V{-}\vn{p})}\right]=0.
\end{multline}
Making the identification of $\alpha_n=\mu_n/T$ and $\beta=-1/T$,
we obtain
\begin{equation}
N_n^k=\frac{G_n^k(V{-}\vn{n})}{1{+}\exp\left[(\varepsilon_n^k-\mut{n})/\kbz
T\right]} \label{Nnk},
\end{equation}
\begin{equation}
 \mut{n}=\mu_n+\kbz
T\sum_{p}\Dif{\vn{p}}{N_n}{}\sum_{i}
G_p^i\ln\left[\frac{G_p^i(V{-}\vn{p}){-}N_p^i}{G_p^i(V{-}\vn{p})}\right].
\label{mu-general}
\end{equation}
We will introduce a new designation $\vnt{n}\equiv\vn{n}/V$ and
will mark the functions pertaining to an ideal Fermi gas by the
superscript $\mathrm{id}$. Summing Eq.~(\ref{Nnk}) over $k$ yields
\begin{equation}
N_n=\left(1{-}\vnt{n}\right)N_n^\id(T,\mut{n}). \label{N_excl_vol}
\end{equation}
Substituting Eq.~$(\ref{Nnk})$ for $N_n^k$ into the double sum of
Eq.~$(\ref{mu-general})$ for the chemical potential, we will find
that
\begin{equation}
\mut{n}=\mu_n-\sum_{p}\Dif{\vnt{p}}{n_n}{}P_{p}^{\id}(T,\mut{p}),
\label{mu-Pressure}
\end{equation}
where $P_{p}^{\id}(T,\mut{p})$ is the pressure of an ideal Fermi
gas as a function of the temperature and chemical potential, while
$n_n$is the number density of the $n$-th mixture component. It is
now easy to derive the expressions for the thermodynamic
potentials:
\begin{equation}
E=\sum_{n}(1{-}\vnt{n})E_{n}^{\id}(T,\mut{n})=\sum_{n}N_n\textbf{E}_{n}^{\id}(T,\mut{n}),
\label{E_excl_vol_general}
\end{equation}
where we introduced the energy per particle
$\textbf{E}_{n}^{\id}=E^{\id}_{n}/N_n$. Similarly,
\begin{align}
S&=\sum_{n}(1{-}\vnt{n})S_{n}^{\id}(T,\mut{n})=\sum_{n}N_n\textbf{S}_{n}^{\id}(T,\mut{n}),\\
F&=\sum_{n}(1{-}\vnt{n})F_{n}^{\id}(T,\mut{n})=\sum_{n}N_n\textbf{F}_{n}^{\id}(T,\mut{n}),
\end{align}
where $F$ is the system’s free energy. We can now determine the
pressure
\begin{equation}
P=\frac{1}{V}\Bigl(\sum_{n}\mu_n N_n{-}F\Bigr)=
\sum_{n}\!\left[1{-}\vnt{n}{+}\!\sum_{i}\Dif{\vnt{n}}{n_i}{}
n_i\right]P_n^\id(T,\mut{n}).\label{P_excl_vol}
\end{equation}
Formulas (\ref{N_excl_vol}){--}(\ref{P_excl_vol}) completely
describe the thermodynamics of matter in terms of the approach to
the EVA under consideration.

It can be shown that this EVA description provides a
thermodynamically consistent description of the system (i.e., an
automatic fulfilment of all thermodynamic relations) for
\emph{any} dependence $\vnt{n}$ on the number densities of the
mixture components.

Thus, the approach considered provides a convenient tool for
investigating the thermodynamics of nonideal matter in terms of
which interaction models can be constructed by choosing the form
of the excluded--volume function $\vnt{n}$ in conformity with the
problem under consideration. For example, let us take an
approximation where the excluded volume is the same for all types
of particles and is just equal to the volume occupied by these
particles: $\vnt{n}=\sum_{i}\textbf{v}_i n_i$, where
$\textbf{v}_i$ is the intrinsic volume of the particles of type
$i$ and $n_i$ is their number density. In this case,
\begin{equation}
P=\sum_{n}P_n^{\id}(T,\mut{n})=P^{\id}(T,\{\mut{n}\}),\quad
\mut{n}=\mu_n{-}\textbf{v}_n P^{\id}(T,\{\mut{n}\}),
\end{equation}
i.e., we obtained the EVA description that was proposed by Rischke
et al. (1991) and that is widely used in describing the reactions
with heavy ions (see, e.g., Hung and Shuryak 1998; Gorenstein et
al. 1996).

\section*{INTERACTION EFFECTS} The above EVA scheme can be supplemented
by the mechanism of allowance for additional interactions between
particles. It is well known that a fairly wide class of
interactions can be described in the formalism of free energy
where the free energy of a system $F(T,V,\{N_i\})$ is represented
as the sum of the system’s free energy without any interaction
$F^{\id}(T,V,\{N_i\})$ and the additional term $\triangle
F(T,V,\{N_i\})$ attributable to the interaction:
$F=F^{\id}{+}\triangle F$. Using this approach, the system’s free
energy in our case should be written as
\begin{equation}
F=\sum_{n}(1{-}\vnt{n})F_{n}^{\id}(T,\mut{n})+\triangle
F(T,V,\{N_i\}).
\end{equation}
Introducing the additional term describing the interaction into
the free energy allows the corresponding changes in the remaining
quantities to be obtained in a thermodynamically consistent way:
\begin{align}
\mut{n}&\Rightarrow\mut{n}+\frac{1}{V}\Dif{\triangle
F}{n_n}{V,T}\label{Dmu_excl_vol}\\
P&\Rightarrow P+\frac{1}{V}\left[\sum_{i}n_i\Dif{\triangle
F}{n_i}{V,T}\!\!-\triangle F\right]\label{DP_excl_vol}\\
S&\Rightarrow S-\Dif{\triangle
F}{T}{V,\{N_i\}}\\
E&\Rightarrow E+\triangle F-T\Dif{\triangle F}{T}{V,\{N_i\}}
\end{align}
These formulas describe the procedure of allowance for additional
interactions in the approach under consideration.

\section*{CONNECTION WITH THE HARD--SPHERE
MODEL} To concretize the form of the excluded--volume function
$\vnt{n}$, let us consider the so-called hard--sphere
approximation, i.e., the model of an ideal gas of particles
interacting through the potential
\begin{equation}
U_{ij}(r)=\left\{
\begin{aligned}
0&,\ \mbox{при}\ r\geq(\sigma_i{+}\sigma_j)/2,\\
{+}\infty&,\ \mbox{при}\  r<(\sigma_i{+}\sigma_j)/2,
\end{aligned}
\right.
\end{equation}
where $r$ is the separation between the mixture components and
$\sigma_i$ are the particle diameters. Numerous analytical results
were obtained in this approximation and its application to the
description of liquid properties was considered. In recent years,
the study has also been carried out using numerical simulations.
Therefore, it is natural to associate our approach in this limit
with the above theory and to try to determine the as yet unknown
quantities. For this purpose, we will set $\triangle F=0$ and will
assume that all particles are Boltzmann ones.

\subsection*{The Single--Component Case} An important parameter of the hard--sphere model
is $\Gamma\equiv P/\left(\kbz T\!\sum_{i}n_i\right)$. Its
deviation from unity characterizes the degree of nonideality of
the system. In the single--component case, this quantity depends
only on the packing fraction $\eta$. Different analytical
approaches give different values for this quantity. The following
expressions are the best known ones:
\begin{equation}
\WID{\Gamma}{PY}=\frac{1{+}2\eta{+}3\eta^{2}}{(1{-}\eta)^{2}},\quad
\WID{\Gamma}{L}=\frac{1{+}\eta{+}\eta^{2}}{(1{-}\eta)^{3}},\quad
\WID{\Gamma}{CS}=\frac{1{+}\eta{+}\eta^{2}{-}\eta^{3}}{(1{-}\eta)^{3}}.
\label{Gamma_one_comp}
\end{equation}
The first two were derived by Lebowitz (1964) based on the
solution of the Percus–Yevick equation for the radial distribution
function; the last was derived by Carnahan and Starling (1969) and
is currently considered more accurate. In the subsequent
applications, we will use precisely this expression.

Let us now consider the expression for the pressure in our
description:
\begin{equation}
P=\left[1{-}\vnt{}{+}n\Dif{\vnt{}}{n}{}\
\right]P^{\id}(T,\mut{})=\left[1{-}\vnt{}{+}n\Dif{\vnt{}}{n}{}\
\right]\frac{nkT}{1{-}\vnt{}},
\end{equation}
whence it follows that
\begin{equation}
1{+}\frac{\eta}{1{-}\vnt{}}\Dif{\vnt{}}{\eta}{}=\Gamma(\eta)\ \
\Rightarrow\ \
\vnt{}(\eta)=1{-}\exp\biggl({-}\!\!\int\limits_{0}^{\eta}\!\frac{\Gamma(x){-}1}{x}\
dx\biggr).
\end{equation}
Thus, in the single-component case, we determined the form of the
dependence of the relative excluded volume $\vnt{}$ on particle
number density that remained unknown. In the limit of low number
densities, $\Gamma(\eta)\approx 1{+}4\eta$ and, hence,
$\vnt{}(\eta)\approx 4\eta$, i.e., the effective excluded volume
is equal to quadruple the intrinsic particle volume, in complete
agreement with the previously reached conclusions.

\subsection*{The Multicomponent Case} It should be noted at once that since there is no
universally accepted approach to describing a multicomponent
system of hard spheres, the approximation presented below is not
the only possible one. On the other hand, this may be considered
as a definite plus, because the breadth of the already described
scheme, when necessary, can accommodate new data on complex
properties of such systems. Our subsequent consideration will be
associated with the work by Lopez de Haro et al. (2008), some of
whose propositions we will use.

For convenience, let us introduce a set of quantities $\psn{n}$
according to the definition: $e^{-\psn{n}}\equiv1{-}\vnt{n}$.
Obviously, $\psn{n}$ are nonnegative. Now, we will make the first
assumption that $\psn{n}=\psn{n}(\eta,\Avsg{k})$, i.e., the number
densities of the components can enter into the expression for
$\psn{n}$ only via the packing fraction $\eta$ and the means
$\Avsg{k}\equiv\sum_{i}\sigma_i^k n_i/\sum_{i}n_i$.

Let us now list the conditions that our model of a multicomponent
system should satisfy:
\begin{enumerate}
\item If the diameter of some component is zero,
$\sigma_k=0$, then these particles must be described by the
formulas for an ideal (in the sense that there is no interaction)
gas, $\mut{k}=\mu_k$, but located in a reduced volume
$V^{*}=V(1{-}\eta)$.\label{COND-Zero}
\item The limit of low number densities described
above in the ``Boltzmann Gas'' Section must hold, i.e.
\begin{equation}
\vnt{n}+\sum_{i}\Dif{\vnt{i}}{n_n}{}\ n_i=
\frac{\pi}{6}\sum_{i}n_i\sigma_{i n}^{3},\quad \sigma_{i
n}\equiv\sigma_i{+}\sigma_n. \label{Low-Formula} \vspace{-1.cm}
\end{equation}\label{COND-Low}
\item If the diameters of all particles are identical,
then all must be reduced to the single-component case
$\Gamma(\{n_i\})=\Gamma_{\mathrm{sc}}(\eta)$. \label{COND-Equal}
\end{enumerate}

These conditions impose moderately stringent constraints on the
form of the function $\psn{n}$, leaving a large room for various
options. Given that in the limit of low number densities
$\psn{n}\approx\vnt{n}$, we will seek $\psn{n}$ in the form
\begin{equation}
\psn{n}=\frac{\pi}{6} \sum_{i}n_i\sgt{ni}{3}G(\eta,z_{n i}),\quad
z_{n i}=2\frac{\sigma_n\sigma_i}{\sigma_{n
i}}\frac{\Avsg{2}}{\Avsg{3}}. \label{PSIn}
\end{equation}
The quantity $\frac{\pi}{6} \sgt{ni}{3}$ means the effective
volume unavailable to the $n$-th particle due to its interaction
with the $i$-th one. The function $G(\eta,z_{n i})\rightarrow 1$
when $\eta\rightarrow 0$. The introduction of $z_{n i}$ is
justified by the comparison with some works on multicomponent
systems (see Lopez de Haro et al. (2008) and references therein).
For our purposes, it will suffice to use the linear expansion
$G(\eta,z_{n i})=g_1(\eta){+}z_{n i}g_2(\eta)$ by assuming that
the term $g_1$ plays a major role at low number densities.
Allowance for additional conditions can require including the next
terms of the expansion in $G(\eta,z_{n i})$ in $z_{n i}$, but we
will restrict ourselves to the linear approximation.

Consider condition \ref{COND-Zero} and let $\sigma_n=0$. The
expressions for the excluded volume and the chemical potential
will be written as
\begin{gather}
1{-}\vnt{n}=1{-}\eta=\exp\!\left({-}\frac{\pi}{6}g_1(\eta)\sum_{i}\sgt{ni}{3}
n_i\right),\\
\mut{n}=\mu_n-\frac{\pi}{6}g_1(\eta)\sum_{i}\sgt{in}{3}n_i.
\end{gather}
Hence and from the meaning of $\sgt{ij}{}$ it follows that
\begin{gather}
\sgt{ni}{}=\sigma_i,\ \mbox{а}\ \ \sgt{in}{}=0 \
\mbox{при}\ \sigma_n=0,\\
g_1(\eta)=-\frac{\ln(1{-}\eta)}{\eta}.
\end{gather}
Let us now pass to condition \ref{COND-Low}. Remembering that
$\WID{g}{1}(\eta)$ plays a major role at low densities, we can
write $\vnt{n}\approx\frac{\pi}{6}\sum_{i}n_i\sgt{ni}{3}$.
Condition $(\ref{Low-Formula})$ can then be rewritten as
\begin{equation}
\sgt{ni}{3}+\sgt{in}{3}=\sigma_{i n}^{3}.
\end{equation}
Now, we should choose the specific form of $\sgt{in}{}$,
satisfying the conditions listed above. We could have used the
expression from Gorenstein et al. (1999), who used the relation
\begin{equation}
\sgt{ni}{3}=\frac{\sigma_i^3\sigma_{n i}^{3}}
{\sigma_i^3{+}\sigma_n^3},
\end{equation}
However, in this case, $\psn{n}$ cannot be represented as
$\psn{n}=\psn{n}(\eta,\Avsg{k})$. Therefore, we propose the
dependence
\begin{equation}
\sgt{ni}{3}=\sigma_i\sigma_{n i}^{2}.
\end{equation}
Note once again that these relations are not the only possible
ones; they are only the simplest ones satisfying the necessary
conditions.

Now, it remains to use requirement \ref{COND-Equal} to find
$g_2(\eta)$. When the diameters of all particles are identical,
$\psn{n}=4\eta\left[g_1(\eta){+}g_2(\eta)\right]$. Substituting
this expression into the formula for $\Gamma$ yields
\begin{equation}
g_2(\eta)=\frac{1}{4\eta}\int\limits_{0}^{\eta}
\left[\WID{\Gamma}{sc}(x){-}\frac{1{+}3x}{1{-}x}\right]\frac{dx}{x},
\end{equation}
where $\WID{\Gamma}{sc}(\eta)$ is the expression for $\Gamma$ in
the single--component case, for example, one of
Eqs.~(\ref{Gamma_one_comp}). Now, it remains to bring all formulas
together and to write explicit expressions for $\psn{n}$ and
$\Gamma$ as a function of $\eta$ and $\Avsg{k}$.
\begin{equation}
\frac{\psn{n}}{\eta}=\left[1{+}2\sigma_n\frac{\Avsg{2}}{\Avsg{3}}
{+}\sigma_n^2\frac{\Avsg{}}{\Avsg{3}}\right] g_1(\eta)+
2\sigma_n\frac{\Avsg{2}}{\Avsg{3}}\left[1{+}\sigma_n
\frac{\Avsg{2}}{\Avsg{3}}\right] g_2(\eta).\label{psn_excl_vol}
\end{equation}
Substituting this expression into the formula for $\Gamma$ and
using the derived expressions for the functions $g_1(\eta)$ and
$g_2(\eta)$, we will obtain
\begin{equation}
\Gamma=\frac{1}{1{-}\eta}+\frac{\Avsg{}\Avsg{2}}{2\Avsg{3}}
\left\{\left[\WID{\Gamma}{sc}(\eta){-}\frac{1}{1{-}\eta}\right]
\left(1{+}\frac{\Avsg{2}^{2}}{\Avsg{}\Avsg{3}}\right)+
\frac{3\eta}{1{-}\eta}\left(1{-}\frac{\Avsg{2}^{2}}
{\Avsg{}\Avsg{3}}\right)\right\}.
\end{equation}
These expressions exactly correspond to the formula for $\Gamma$,
derived in the same approximation by Lopez de Haro et al. (2008).

Thus, using Eq.~(\ref{psn_excl_vol}), we were able to relate
$\psn{n}$ and, via it $\vnt{n}$ to the number densities and
diameters of the particles constituting the system. However, our
formulas are applicable not only for an ideal gas but also for
degenerate particles and when additional interactions are present
in the system.

\section*{MATTER IN THE SUBNUCLEAR RANGE} To derive the equation of state in the subnuclear
range that provides the phase transition to uniform nuclear matter
at densities $\rho\sim 10^{14}\ \gccm$, it is necessary to add the
long-range attractive interaction potential to the short--range
repulsive potential described in our approach by the
excluded-volume effect. The well--known Yukawa potential
$V(r)\propto e^{-\lambda r}/r$, where $\lambda$ is determined by
the pion Compton wavelength $l_{\pi}=1/\lambda\approx
1.41~\mbox{fm}$, is used to describe the interaction of isolated
nucleons at great distances. In reality, the interaction between
nucleons in the range of nuclear densities is not reduced to the
sum of pair interactions. This forces one to introduce the concept
of the so-called effective interaction potential whose form and
parameters are chosen in such a way as to reproduce the properties
of uniform nuclear matter, such as the saturation density
$n_{\mathrm b}\approx 0.16\ \mbox{fm}^{-3}$, the binding energy
per nucleon $B\approx -16~\mbox{MeV}$ etc. For example, the
effective Seyler–-Blanchard potential is the product of a Yukawa
potential with a parameter $1/\lambda\approx 0.6\ \mbox{fm}$ by a
function dependent both on the local matter density and on the
momenta of the interacting nucleons (for more detail, see Myers
and Swiatecki 1969).

We will use a Yukawa potential whose parameters are chosen from
the condition for consistency with the equation of state for
nuclear matter. This form of the potential was chosen only because
of its simplicity in order to demonstrate the efficiency of the
approach being described. As we will see below, in general, the
results are rather sensitive to the parameters of the potential
and choosing it to obtain realistic results is a nontrivial
problem.

The additional complexity is that we should include the
interactions between free nucleons and between nuclei in the
description. This question is considered in the Appendix, where
the specific values of parameters used in our calculations are
also given.

\section*{RESULTS AND DISCUSSION}
We will apply the developed approach to describe the subnuclear
density range for supernova matter under NSE conditions. In the
subsequent figures, unless specified otherwise, the temperature
$T=5~\mbox{MeV}$ and $\Ye=1/3$ everywhere. The results of the
model including only the excluded-volume effect are marked as
``EV'', while those of the model that, apart from this effect,
includes the long-range potential are marked as ``EV{+}LRA''.

\begin{figure}[h!]
\begin{center}
\vspace*{-0.3cm}
\includegraphics[totalheight=8cm]{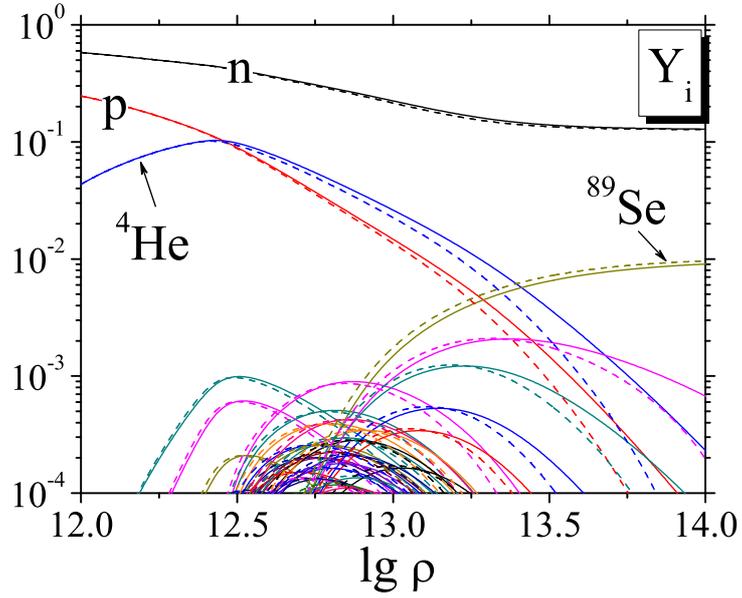}
 \vspace*{-0.2cm}
 \caption{(Color online) Chemical composition of matter at $T=5~\mbox{MeV}$ and $\Ye=1/3$, the EV model.}
\label{PicYi_compare}
\vspace*{-0.3cm}
\end{center}
\end{figure}

\begin{figure}[h!]
\begin{center}
\includegraphics[totalheight=8cm]{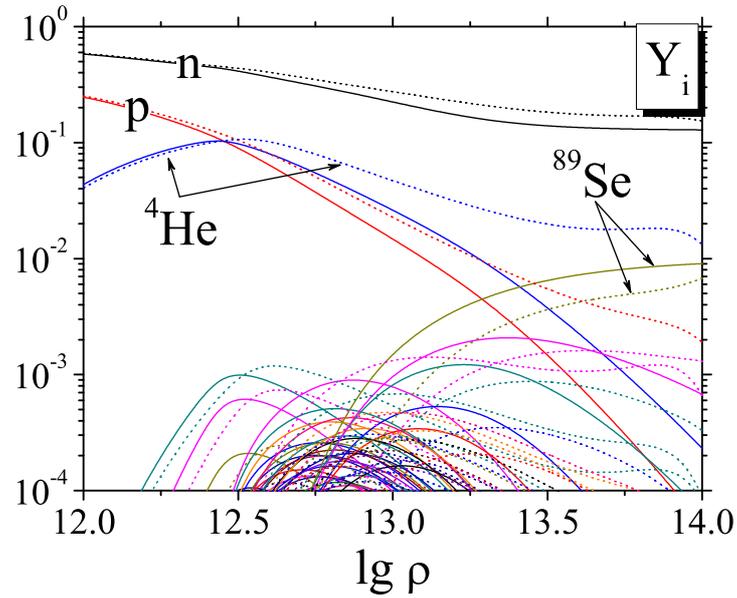}
 \vspace*{-0.2cm}
 \caption{(Color online) Chemical composition of matter at $T=5~\mbox{MeV}$ and $\Ye=1/3$, model $\mathrm{EV}{+}\mathrm{LRA}$}
\label{PicYi_compareLRA}
 \vspace*{-0.3cm}
\end{center}
\end{figure}
Figure $\ref{PicYi_compare}$ shows the equilibrium chemical
composition of matter as a function of the density. The solid
lines represent the calculation with ideal matter; the dashed
lines indicate the result of the EV model. Similar data are
presented in Figure~$\ref{PicYi_compareLRA}$, with the only
difference that the dotted lines indicate the result of our
calculation according to the EV{+}LRA model.

\begin{figure}[htb]
\begin{center}
\vspace*{-0.3cm}
\includegraphics[totalheight=8cm]{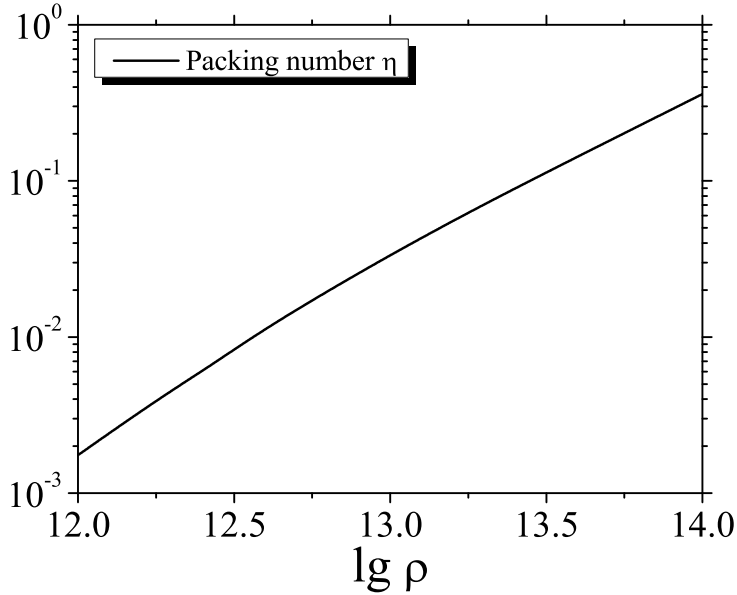}
 \vspace*{-0.2cm}
 \caption{Packing fraction $\eta$}
\label{PicPackingNumber}
 \vspace*{-0.3cm}
\end{center}
\end{figure}
Figure \ref{PicPackingNumber} shows the calculated packing
fraction $\eta$ under the same conditions. As we see, the
excluded--volume effect per se has a weak influence on the
equilibrium mass fractions up to $\rho=10^{14}~\rgccm$, although
the packing fraction reaches a significant value here,
$\eta\approx 0.4$. Allowance for the long--range part of the
potential gives a considerably stronger effect: the number density
of free nucleons and $\alpha$-particles increases; the number
density of heavy nuclei, in particular, the neutron-rich selenium
isotopes that are representatives of all neutron-rich nuclei in
our set of nuclides, decreases (for more detail, see Nadyozhin and
Yudin 2004). Thus, first, the long--range component of the
effective nucleon–nucleon interaction potential itself and,
second, the method for generalizing this interaction to nuclei are
important for a proper description of the subnuclear range. the
widely used mean--nucleus models, a decrease in the surface energy
of the nucleus due to the medium’s effects corresponds to this
interaction. Our method is presented in the Appendix.

\begin{figure}[htb]
\begin{center}
\vspace*{-0.3cm}
\includegraphics[totalheight=8cm]{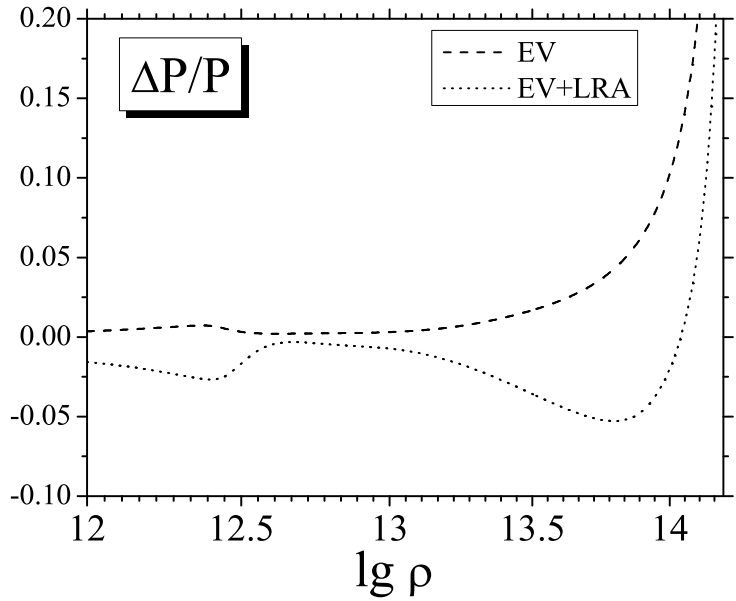}
 \vspace*{-0.2cm}
 \caption{Relative change in the pressure of matter.}
\label{PicRelPressure}
 \vspace*{-0.3cm}
\end{center}
\end{figure}
Figure $\ref{PicRelPressure}$ shows the relative change in
pressure $(P{-}P^{\mathrm{id}})/P^{\mathrm{id}}$, where
$P^{\mathrm{id}}$ is the pressure of ideal matter. The dashed and
dotted lines indicate the EV and EV{+}LRA models, respectively.
The non--monotonic behavior of the curves at $\lg\rho\approx 12.5$
is associated with the change in equilibrium chemical composition
under the influence of interaction. At $\rho\sim 10^{14}~\rgccm$
the pressure in both models rises sharply.

\begin{figure}[h!]
\begin{center}
\vspace*{-0.1cm}
\includegraphics[totalheight=8cm]{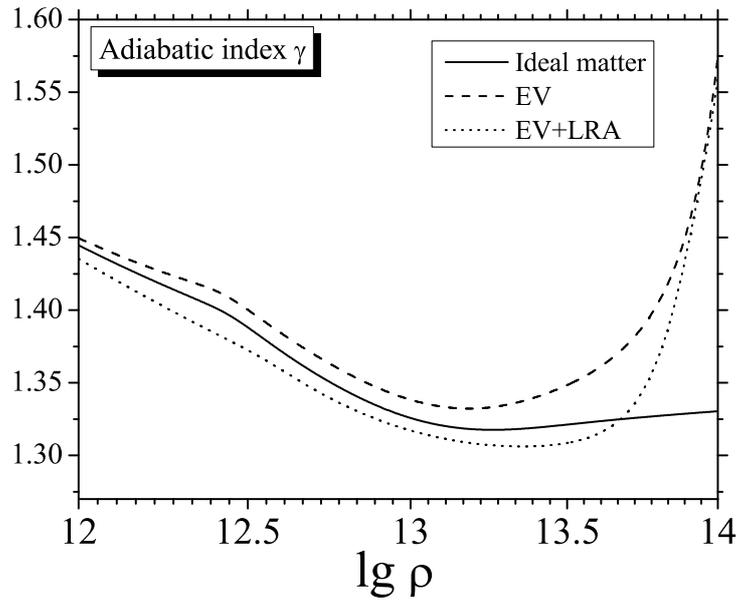}
 \vspace*{-0.2cm}
 \caption{Adiabatic indices for matter.}
\label{PicGamma}
 \vspace*{-0.3cm}
\end{center}
\end{figure}
Figure $\ref{PicGamma}$ shows the behavior of the adiabatic index
$\gamma\equiv\left(\partial\ln P/\partial\ln\rho\right)_S$ for
ideal matter (solid line), the EV model (dashed line), and the
EV{+}LRA model (dotted line). As would be expected, the EV model
makes the matter ``stiffer'' in the entire density range, while
the long--range potential of the EV{+}LRA model causes $\gamma$ to
be reduced in the range of low densities. At $\rho\gtrsim 10^{14}$
the equation of state in the excluded-volume models becomes very
``stiff''. The phase transition to uniform nuclear matter to be
discussed in the next section should occur in this range.

\subsection*{The Phase Transition}
Naturally, only the equation of state defined in the entire domain
of thermodynamic parameters of interest to us and containing the
low-- and high--density phases as their limiting cases can give an
absolutely proper description of the phase transition. The EVA is
a low--density approximation of the real equation of state and
cannot be applied at excessively high densities. However, as we
will show below, the EVA can be used in this capacity in the phase
equilibrium equations (equality between the pressures and chemical
potentials of the phases) provided that the equation of state for
uniform nuclear matter (as such we use the results of the approach
by Lattimer and Swesty (1991)) is used for the high--density
phase. In this case, the parameters of both phases should be
reconciled, which is a separate nontrivial problem. In fact, the
procedure described above provides a thermodynamically proper
joining of the two equations of state. In particular, this
approach proved to be good in our hydrodynamic simulations of
gravitational collapse.

\begin{figure}[hbt]
\begin{center}
\vspace*{-0.2cm}
\includegraphics[totalheight=8cm]{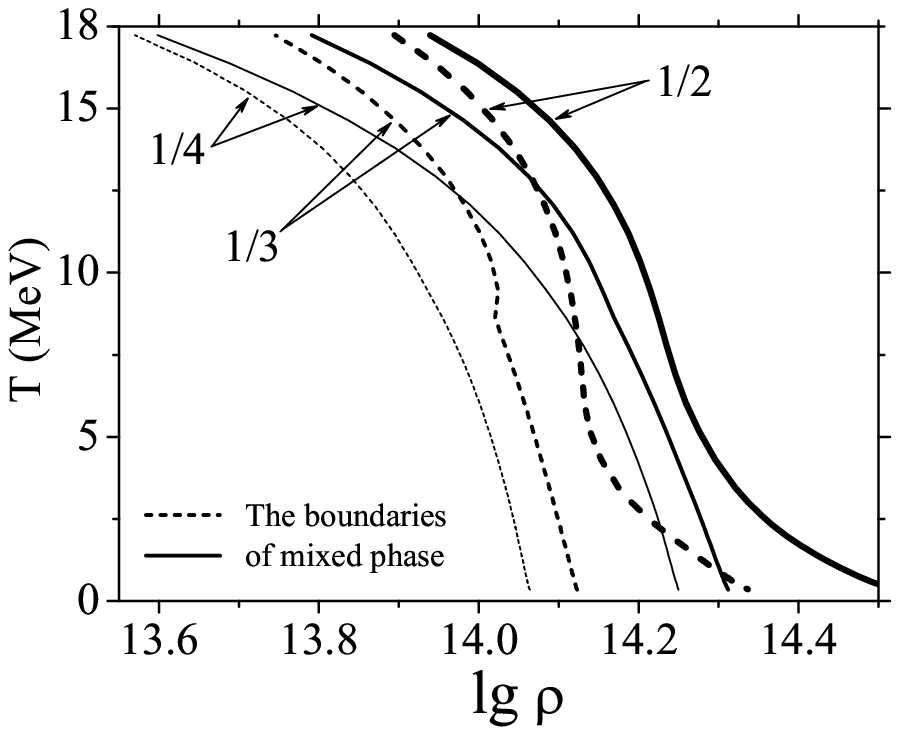}
 \vspace*{-0.2cm}
 \caption{Фазовая диаграмма вещества}
\label{PicPhaseDiagram}
 \vspace*{-0.3cm}
\end{center}
\end{figure}
Figure $\ref{PicPhaseDiagram}$ shows the phase diagram for matter
calculated using the procedure described above with the EV{+}LRA
model for the equation of state for the low--density phase. The
dashed lines indicate the boundary between the low-density phase
and the region of mixed states; the solid lines indicate the
boundary between the mixed states and the high--density phase
(uniform nuclear matter). The numbers with arrows indicate the
leptonic charges $\Ye$ for each pair of lines: the thick, thinner,
and thinnest lines correspond to $\Ye=1/2$, $\Ye=1/3$ and
$\Ye=1/4$ respectively. As the temperature rises, the lines of the
phase boundaries converge, i.e., the equations of state for the
phases become increasingly close. On the real phase diagram for
matter, there must be a critical point with a certain temperature
$T_\mathrm{c}$ above which the phases are indistinguishable in
this region.

\begin{figure}[htb]
\begin{center}
\vspace*{-0.3cm}
\includegraphics[totalheight=8cm]{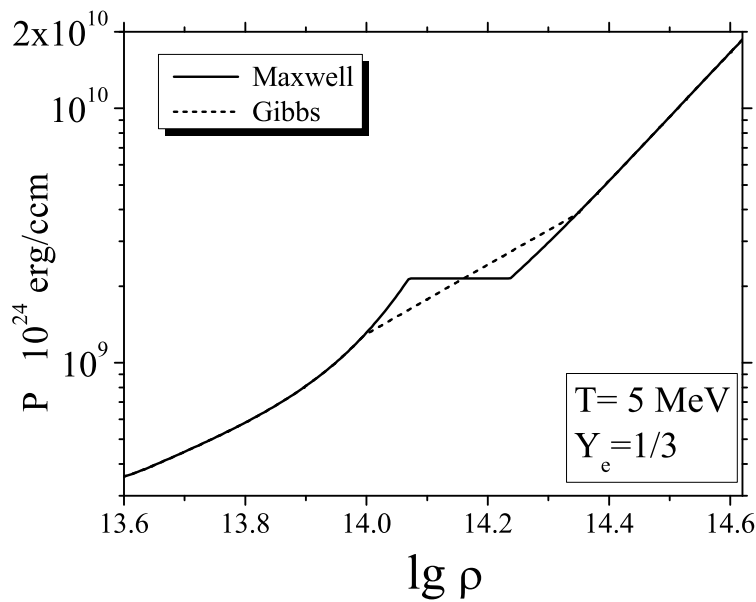}
 \vspace*{-0.2cm}
 \caption{Pressure during the phase transition (see the text).}
\label{PicMaxGibbs}
 \vspace*{-0.3cm}
\end{center}
\end{figure}
Figure $\ref{PicMaxGibbs}$ shows the behavior of the matter
pressure during the phase transition. The solid line with a
characteristic plateau $P=\mathrm{const}$ corresponds to the
ordinary description of the phase transition according to
Maxwell’s approach. The latter requires that the pressures and
chemical potentials of the phases be equal in the domain of their
existence, with each phase being considered electrically neutral.
Gibbs’s approach, whose result is indicated by the dotted line,
allows the phases to have an uncompensated charge, providing only
global electrical neutrality. As we see, $\partial
P/\partial\rho>0$, in Gibbs’s approach in the region of mixed
states, but this region itself is wider. The phase diagram
presented above was calculated in accordance with Maxwell’s
ordinary approach.

\section*{CONCLUSIONS}
The general approach to the EVA developed here can serve as a tool
for investigating the extreme states of matter. Different
thermodynamically consistent models for the equation of state can
be obtained by choosing different forms of the excluded--volume
function $\vnt{n}$ and the corresponding additional interaction
potential.

Using this EVA method, it turned out to be possible to reproduce
the results of the hard--sphere model in the Boltzmann limit. This
approach is apparently adequate for describing a multicomponent
mixture of free nucleons and nuclei under NSE conditions, i.e.,
the case of supernova matter at subnuclear densities. It can be
used not only to study the thermodynamic properties of matter but
also to obtain detailed information about its chemical
composition. This is a serious advantage of our approach over the
popular mean-nucleus models. For example, the nucleosynthesis
problems can be solved by using only this type of equations of
state.

In addition, we showed that based on this approach to the EVA, we
can obtain the phase transition to uniform nuclear matter and,
hence, use this equation of state in hydrodynamic simulations of
supernova explosions.

\vspace*{1.cm}
\appendix

\LARGE APPENDIX \normalsize \subsection*{THE INTERACTION OF
NUCLEI}
 Let the interaction energy between two nucleons
of a certain type, for example, an $n{-}p$, separated by a
distance $r$ be $\WID{V}{Y}(r)=Y e^{-\lambda r}/\lambda r$. The
interaction energy between one nucleon $(n)$ and nucleons of a
given type $(p)$ in the nucleus will then be
\begin{equation*}
E_{\mathrm{int}}=\int\WID{V}{Y}(r)\rho(r)d^{3}r,\quad
N=\int\limits_{0}^{R}4\pi r^{2}\rho(r)dr,
\end{equation*}
where $N$ is the number of nucleons of the type under
consideration in the nucleus, $R$ is its radius, $\rho(r)$ is the
density of the distribution of nucleons in the nucleus, and the
distribution itself is assumed to be spherically symmetric. The
integration in the formula for $E_{\mathrm{int}}$ is over the
entire nucleus volume. We obtain
\begin{equation*}
E_{\mathrm{int}}=\int\limits_{0}^{R}\!\!
\int\limits_{0}^{\pi}\WID{V}{Y}(l)2\pi r^{2}\rho(r)dr\sin\theta
d\theta=N\WID{V}{Y}(L)G_{\rho}(\lambda,R).
\end{equation*}
Here, $l=\sqrt{L^{2}{+}r^{2}{-}2Lr\cos\theta}$, $L$ is the
distance between the nucleon and the nucleus center, and
\begin{equation*}
G_{\rho}(\lambda,R)=\frac{1}{\lambda}\left[\int\limits_{0}^{R}\rho(r)r\sinh(\lambda
r)dr\right]\left[\int\limits_{0}^{R}\rho(r)r^{2}dr\right]^{-1}
\end{equation*}
Thus, the interaction between a nucleon and a uniformly
``charged'' nucleus is described by the same Yukawa potential,
only with the corrected factor $G_{\rho}(\lambda,R)$. Here, there
is an analogy with the interaction via a Coulomb (or
gravitational) potential, where a uniformly charged sphere is
equivalent to an equal (in magnitude) point charge placed at its
center (see also Azam and Gowda 2005). Below, for simplicity, we
will assume that the density of nucleons in the nucleus is
constant. As a result, $G_{\rho}(\lambda,R)$ transforms into
$G(\lambda R)$, where
\begin{equation*}
G(x)=\frac{3}{x^{2}}\left[\cosh(x)-\frac{\sinh(x)}{x}\right],\quad
G(0)=1.
\end{equation*}
Since the potential has the Yukawa form as before, it is not
difficult to write out the interaction energy between two nuclei
separated by a distance $l$ in the final form; it is only
necessary to take into account the fact that the $n{-}n$ and
$p{-}p$ pairs interact identically, while the interaction of the
$p{-}n$ differs from them:
$\WID{Y}{nn}=\WID{Y}{pp}\neq\WID{Y}{np}$.
\begin{align*}
\WID{E}{int}&=\frac{e^{-\lambda l}}{\lambda l}P_{1,2}G(\lambda
R_{1})G(\lambda R_{2}),\\
P_{12}&=\WID{Y}{nn}(N_{1}N_{2}+Z_{1}Z_{2})+\WID{Y}{np}(N_{1}Z_{2}+N_{2}Z_{1}),
\end{align*}
where $N$ and $Z$ are the numbers of neutrons and protons in the
nucleus, respectively. Now, it is easy to write the expression for
the total energy per unit volume of the (free nucleons + nuclei)
system:
\begin{multline*}
\WID{E}{tot}=\frac{1}{2}\sum\limits_{i,j}n_{i}n_{j}\!\!\int\limits_{R_{i}+R_{j}}^{\infty}\!\!
\WID{E}{int}d^{3}L=\\
=\frac{2\pi}{\lambda^{3}}\sum\limits_{i,j}n_{i}n_{j}P_{ij}\
G(\lambda R_{i})G(\lambda
R_{j})e^{-\lambda(R_{i}+R_{j})}[1{+}\lambda(R_{i}{+}R_{j}))],
\end{multline*}
where $n_{i}$ and $n_{j}$ are the number densities of the
components. It is easy to determine $\WID{Y}{nn}{=}\WID{Y}{pp}$
and $\WID{Y}{np}$ by comparing this expression with the formula
for the energy of uniform nuclear matter (see Lattimer and Swesty
1991). We take the parameter $\lambda$ to be $0.95\ \mbox{fm}$ and
the radii of the nuclei and nucleons to be $\WID{R}{A,Z}=1.16
A^{1/3}\ \mbox{fm}$ and $\WID{R}{n}=\WID{R}{p}=0.8\ \mbox{fm}$,
respectively.

\vspace*{1.cm}
\appendix

\large ACKNOWLEDGMENTS \normalsize
 \vspace*{0.5cm}

I wish to thank D.K.~Nadyozhin for numerous helpful discussions of
the questions considered. This work was supported by an SNSF grant
(SCOPES project IZ73Z0-128180/1) and the Russian Federal Agency
for Science and Innovations (project 02.740.11.0250).


\end{document}